\documentclass[10pt]{article}
\usepackage{latexsym}
\usepackage{amssymb}
\usepackage{amsmath}
\usepackage{amscd}
\usepackage{amsthm}
\usepackage[left=2cm,top=2.5cm,right=2.5cm,bottom=1.5cm]{geometry}
\usepackage[dvips]{graphicx}
\usepackage{hyperref}
\begin{document}
\begin{center}
\large{\bf{The Bulk Viscous String Cosmology in An Anisotropic
Universe With Late Time Acceleration
}} \\
\vspace{10mm}
\normalsize{Hassan Amirhashchi}\\
\vspace{5mm}
\normalsize{Department of Physics, Mahshahr Branch, Islamic Azad University,  Mahshahr, Iran \\
E-mail:h.amirhashchi@mahshahriau.ac.ir} \\
\end{center}

\begin{abstract}
A model of a cloud formed by massive strings is used as a source of Bianchi type II.
We assumed that the expansion $(\theta)$ in the model is proportional to the shear $(\sigma)$.
To get exact solution, we have considered the equation of state of the fluid to be in the stiff form.
It is found that the bulk viscosity plays a very important rule in the history of the universe. In presence of bulk viscosity
the particles dominate over strings whereas in absence of it, strings dominate over the particles which is not in consistence
with the recent observations. Also we observe that the viscosity caused the expansion of the universe to be accelerating.
 Our models are evolving from an early decelerating phase to a late time accelerating phase.
The physical and geometrical behavior of these models are discussed.\\\\
keywords: LRS Bianchi type II models, Stiff Fluid, Massive String\\
PACS: 98.80.-k, 98.80.Cq, 04.20.-q, 04.20.Jb
\end{abstract}
\section{Introduction}
\label{intro} In recent years, there has been considerable
interest in the study of the role of cosmic strings in the evolution of the universe, specially at it's early epoches. One of the predictions of Grand
Unified Theories (GUT) is that the universe underwent a phase
transition as the temperature falls down below the
$T_{GUT}\backsim 10^{28}K$  when the age of universe was
$t_{GUT}\backsim 10^{-36}s$ (Zel'dovich et al. 1975; Kibble. 1980;
Everett. 1981; Vilenkin. 1981). There is a loss of symmetry when
the universe undergoes the GUT phase transition at $t_{GUT}$. At
$T<T_{GUT}$, the symmetry between the strong and electro-weak
forces spontaneously broken. The phase transitions associated with
loss of symmetry leads to the formation of topological defects
such as domain walls, cosmic strings, monopoles, etc. As mentioned by Kibble (1976), cosmic strings which are
important topologically stable defects, might
be created during a phase transition in the early universe. Although at the present
time the existence of the cosmic strings can not be detected through our observations, their existence in the early epoches of cosmic evolutions is a well established fact. Zel'dovich (1980) was believed that the
vacuum strings may generate density fluctuations sufficient to
explain the galaxy formation. Since the cosmic
strings have coupled stress-energy to the gravitational field,
the study of gravitational effects of such strings seems to
be interesting. The general relativistic treatment of strings was
initiated by Letelier (1979, 1983). Here we have considered
gravitational effects, arisen from strings by coupling of stress
energy of strings to the gravitational field. Letelier (1979)
defined the massive strings as the geometric strings
(massless) with particles attached along its expansions.\\
As noted by Belinchon (2009), the string that form the cloud are the generalization of Takabayasi's relativistic model of strings (called $p$-strings).
Therefore, a cloud model of strings is a model in which we have particles and strings together. However, since there is not any observational evidence for the exitance of the cosmic strings at the present time, one can eliminate the strings and end
up with a cloud of particles at the present time of evolution
of the universe (Banerjee et al. 1990; Yadav et al. 2007; Saha and Visinescu. 2008, 2010).\\

Most cosmological models assume that the matter in the universe
can be describe by 'dust' (a pressure-less distribution) or at
best a perfect fluid. To have realistic cosmological models we
should consider the presence of a material distribution other than
a perfect fluid. Cosmological models of a fluid with viscosity
play a significant role in the study of evolution of universe. The
viscosity mechanism in cosmology can account for high entropy per
baryon of the present universe (Weinberg. 1972). It is well known
that at an early stage of the universe when neutrino decoupling
occurs, the matter behaves like a viscous fluid (Kolb and Turner.
1990). Weinberg (1971, 1972) derived general formulae for bulk and
shear viscosity and used these to evaluate the rate of
cosmological entropy production. He deduced that the most general
form of the energy-momentum tensor allowed by rotational and
space-inversion invariance, contains a bulk viscosity term
proportional to the volume expansion of the model. Padmanbhan and
Chitre (1987) have also noted that viscosity may be of relevance
for the future evolution of the Universe. If the coefficient of
bulk viscosity decays sufficiently slowly, i.e.,
$\xi\sim\rho^{n},~n<\frac{1}{2}$, then the late epochs of the
Universe will be viscosity dominated, and the Universe will enter
a final inflationary era with steady-state character. Cosmological
models with viscous fluid in early universe have been widely
discussed in the literatures (for example see Pradhan et al. 2012;
Pradhan and Lata. 2011; Pradhan. 2009; Pradhan and Shyam. 2009;
Pradhan et al. 2008; Pradhan et al 2007; Pradhan et al. 2005
Yadav. 2011 a, b; Yadav et al. 2012
).\\

The equation of state of a cosmic fluid is defined as the ratio of it's pressure and energy density i.e $\omega=\frac{p}{\rho}$.
A fluid in which $p=\rho$, is called ``stiff fluid". In this case, the speed of light is equal to speed of
sound and its governing equations have the same characteristics as
those of gravitational field (Zel'dovich. 1970). The relevance of stiff equation of state $\rho=p$ to the
matter content of the universe in the early state of evolution of
universe has first discussed by Barrow (1986). An exact solution of
Einstein's field equation with stiff equation of state has investigated by Wesson (1978). Mohanty et
al. (1982) have investigated cylindrically symmetric Zel'dovich
fluid distribution in General Relativity. G\"{o}tz (1988) obtained
a plane symmetric solution of Einstein's field equation for stiff
perfect fluid distribution. Pradhan and Kumhar (2009) have
investigated LRS Bianchi type II bulk viscous universe with
decaying vacuum energy density in General Relativity. Recently A. K. Yadav et al. (2011) have investigated string L.R.S Bianchi type II universe in general relativity.\\

Bianchi type II space-time has a fundamental role in constructing
cosmological models suitable for describing the early stages of
evolution of universe. Asseo and Sol (1987) emphasized the
importance of Bianchi type II universe. In the  present  paper we
have considered  a  locally  rotationally  symmetric (LRS) model
of spatially  homogeneous  Bianchi  type-II  cosmology. To obtain
exact  solutions, the field equations have been solved for the
case when the equation of state of the fluid is in the stiff form.
The paper is organized as follows. The metric and the field
equations are presented in Section 2. In Section 3, we deal with
solution of the field equations with cloud of strings. In Subsect.
3.1, we describe some physical and geometric properties of the
model . In Subsect. 3.2, we give the solution in absence of bulk
viscosity. A dark energy interpretation of the derived models is
given in Section 4. Finally, in Section 5, concluding remarks are
given.
\section{The Metric and Field Equations}
\label{sec:1}
We consider the Bianchi type II metric in the form
\begin{equation}
\label{eq1} ds^{2}=-dt^{2}+B^{2}(dy+xdz)^{2}+A^{2}(dx^{2}+dz^{2})
\end{equation}
where $A$ , $B$ are functions of $t$ only. The energy-momentum tensor for a cloud of strings in presence of bulk viscosity is taken as
\begin{equation}
\label{eq2} T_{ij}=(p+\rho)u_{i}u_{j}+pg_{ij}-\lambda x_{i}x_{j}+\xi\theta(g_{ij}+u_{i}u_{j}),
\end{equation}
where $u_{i}$ and $x_{i}$ satisfy condition
\begin{equation}
\label{eq3} u_{i}u^{i}=-x^{i}x_{i}=-1,\quad u^{i}x_{i}=0,
\end{equation}
$p$ is the isotropic pressure, $\rho$ is the proper energy density for a cloud
string with particles attached to them, $\lambda$ is the string tension density, $u^{i}$ the
four-velocity of the particles, and $x^{i}$ is a unit space-like vector representing
the direction of string. In a co-moving coordinate system, we have
\begin{equation}
\label{eq4} u^{i}=(0,0,0,1),\quad x^{i}=(\frac{1}{B},0,0,0).
\end{equation}
The particle density of the configuration is given by
\begin{equation}
\label{eq5} \rho_{p}=R_{ij}u^{i}u^{j},
\end{equation}
where $ \rho_{p}$ is the rest energy density of the particles attached to the strings. The
string tension density,$\lambda$, can take positive or negative values. Negative value
of $\lambda$ represents a universe filled with no strings but only an anisotropic fluid
whereas its positive value represents strings loaded with particles forming the
surface of world sheet \cite{ref8}.\\

The Einstein's field equations (with $8\pi G=1$ and $c=1$)
\begin{equation}
\label{eq6} R_{ij}-\frac{1}{2}Rg_{ij}=-T_{ij}
\end{equation}
for the metric (\ref{eq1}) leads to the following system of equations:
\begin{equation}
\label{eq7} G_{22}=2\frac{\ddot{A}}{A}+\frac{\dot{A}^{2}}{A^{2}}-\frac{3}{4}\frac{B^{2}}{A^{4}}=-p+\xi\theta+\lambda
\end{equation}
\begin{equation}
\label{eq8} G_{11}=G_{33}=\frac{\ddot{A}}{A}+\frac{\ddot{B}}{B}+\frac{\dot{A}\dot{B}}{AB}+\frac{1}{4}\frac{B^{2}}{A^{4}}=-p+\xi\theta
\end{equation}
\begin{equation}
\label{eq9}G_{00}=2\frac{\dot{A}\dot{B}}{AB}+\frac{\dot{A}^{2}}{A^{2}}-\frac{1}{4}\frac{B^{2}}{A^{4}}=\rho
\end{equation}
where an overdot stands for the first and double overdot for second derivative with respect to $t$.\\
The spatial volume for LRS B-II is given by
\begin{equation}
\label{eq10} V = A^{2}B.
\end{equation}
We define $S = (A^{2}B)^{\frac{1}{3}}$ as the average scale factor of LRS B-II model (\ref{eq1})
so that the Hubble's parameter is given by
\begin{equation}
\label{eq11} H = \frac{\dot{S}}{S} = \frac{1}{3}\left(\frac{2\dot{A}}{A} + \frac{\dot{B}}{B}\right).
\end{equation}
We define the generalized mean Hubble's parameter H as
\begin{equation}
\label{eq12} H = \frac{1}{3}(H_{x} + H_{y} + H_{z}),
\end{equation}
where $H_{x} = \frac{\dot{A}}{A}$, $H_{y} = \frac{\dot{B}}{B}$ and $H_{z} = H_{x}$  are
the directional Hubble's parameters in the directions of $x$, $y$ and $z$ respectively. \\\\
The deceleration parameter $q$ is conventionally defined by
\begin{equation}
\label{eq13} q = - \frac{S\ddot{S}}{\dot{S}^{2}}.
\end{equation}
The scalar expansion $\theta$, shear scalar $\sigma^{2}$ and the average anisotropy parameter $Am$
are defined by
\begin{equation}
\label{eq14}
\theta = \frac{2\dot{A}}{A} + \frac{\dot{B}}{B},
\end{equation}
\begin{equation}
\label{eq15}
\sigma^{2} = \frac{1}{2}\left(\sum_{i=1}^{3}H_{i}^{2}-\frac{1}{3}\theta^{2}\right),
\end{equation}
\begin{equation}
\label{eq16} A_{m} = \frac{1}{3}\sum_{i = 1}^{3}{\left(\frac{\triangle
H_{i}}{H}\right)^{2}},
\end{equation}
where $\triangle H_{i} = H_{i} - H (i = 1, 2, 3).$\\\\
The Raychaudhuri equation reads
\begin{equation}
\label{eq17}
3\frac{\ddot{S}}{S}=-2\sigma^{2}+\frac{3}{2}\xi\theta-\frac{1}{2}(\rho+3p).
\end{equation}
\section{Solution of the Field Equations}
The field equations (\ref{eq11})-(\ref{eq13}) are a system of three equations with five unknown parameters $A, B, p, \rho, \lambda$.
two additional constraints relating these parameters are required to obtain explicit solutions of the system. Firstly, we assume that the expansion $\theta$ in the model is proportional to the shear $\sigma$. This condition leads to
\begin{equation}
\label{eq18}\frac{1}{\sqrt{3}}\left(\frac{\dot{A}}{A}-\frac{\dot{B}}{B}\right)=b\left(\frac{\dot{A}}{A}+2\frac{\dot{B}}{B}\right)
\end{equation}
which yields to
\begin{equation}
\label{eq19}\frac{\dot{A}}{A}=m\frac{\dot{B}}{B}
\end{equation}
where $m=\frac{2b\sqrt{3}+1}{1-b\sqrt{3}}$ and $b$ are constants. Eq. (\ref{eq19}), after integration, reduces to
\begin{equation}
\label{eq20} A=B^{m},
\end{equation}

Secondly, we assume that the fluid obeys the stiff fluid equation of state i.e
\begin{equation}
\label{eq21} p=\rho.
\end{equation}
Using eqs. (\ref{eq8}) and (\ref{eq9}) in eq. (\ref{eq21}) we get
\begin{equation}
\label{eq22}\frac{\ddot{A}}{A}+\frac{\ddot{B}}{B}+3\frac{\dot{A}\dot{B}}{AB}+\frac{\dot{A}^{2}}{A^{2}}-\xi\theta=0.
\end{equation}
In view of eq. (\ref{eq20}), eq. (\ref{eq21}) is taken as
\begin{equation}
\label{eq23} 2\ddot{B}+4m\frac{\dot{B}^{2}}{B}=\frac{\chi}{(m+1)}B.
\end{equation}
where I have assumed that the coefficient of bulk viscosity is inversely proportional to expansion i.e $\xi\theta=\chi$ (say) = constant.\\
Let $\dot{B}=f(B)$ which implies that $\ddot{B}=ff'$, where $f'=\frac{df}{dB}$. Hence (\ref{eq23}) can be written as
\begin{equation}
\label{eq24}\frac{d}{dB}(f^{2})+4m\frac{f^{2}}{B}=\frac{\chi}{(m+1)}B,
\end{equation}
which on integration gives
\begin{equation}
\label{eq25}dt=\frac{dB}{\sqrt{aB^{2}+bB^{-4m}}}.
\end{equation}
Here, $a=\frac{\chi}{(2m^{2}+3m+1)}$ and $b$ is a positive integrating constant.
Hence the model (\ref{eq1}) is reduced to
\begin{equation}
\label{eq26}ds^{2}=-\frac{dB^{2}}{aB^{2}+bB^{-4m}}+B^{2}(dx+zdy)^{2}+B^{2m}(dy^{2}+dz^{2}).
\end{equation}
After using a suitable transformation of coordinates the model (\ref{eq26}) reduces to
\begin{equation}
\label{eq27}ds^{2}=-\frac{dT^{2}}{aT^{2}+bT^{-4m}}+T^{2}(dx+zdy)^{2}+T^{2m}(dy^{2}+dz^{2}).
\end{equation}
\subsection{The Geometric and Physical Significance of Model}
Here we discuss some physical and kinematic properties of string model (\ref{eq27}).\\ The pressure $(p)$, the energy density $(\rho)$, the string tension $(\lambda)$, and the particle density $(\rho_{p})$ for the model (\ref{eq27}) are given by
\begin{equation}
\label{eq28} p=\rho=\frac{m(m+2)}{2}\left[\frac{\chi}{(2m^{2}+3m+1)}+2bT^{-2(1+2m)}\right]-\frac{1}{4}\frac{1}{T^{2(2m-1)}},
\end{equation}
\begin{equation}
\label{eq29} \lambda=\frac{1}{2}\left(\frac{2m^{2}-3m+1}{2m^{2}+3m+1}\right)\chi-\frac{1}{T^{2(2m-1)}},
\end{equation}
\begin{equation}
\label{eq30}\rho_{p}=-\frac{1}{2}\left(\frac{m^{2}-5m+1}{2m^{2}+3m+1}\right)\chi+m(m+2)bT^{-2(1+2m)}+\frac{3}{4}\frac{1}{T^{2(2m-1)}}.
\end{equation}
\begin{figure}[htbp]
\centering
\includegraphics[width=8cm,height=8cm,angle=0]{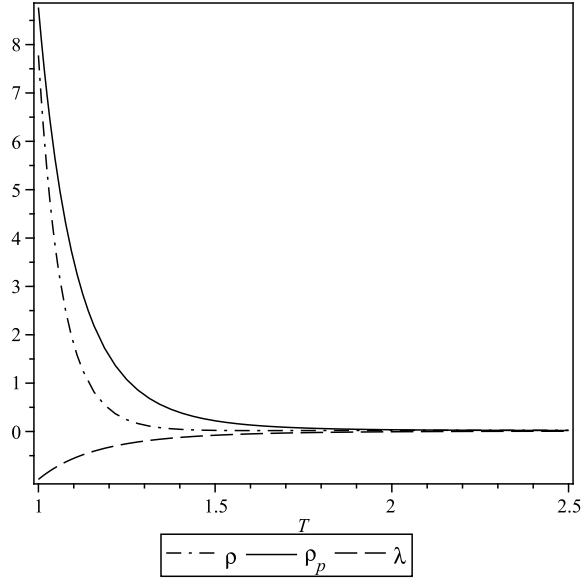}
\caption{The plot of energy density $\rho$, particle density $\rho_{p}$ and tension density $\lambda$ Vs. $T$  for $m=2$, $\chi=0.1$ and $b=1$ }
\end{figure}
From Eqs. (\ref{eq28}) and (\ref{eq30}), we observe that the energy density $\rho$ and the particle density $\rho_{p}$ are decreasing functions of time. This behavior of $\rho$ and $\rho_{p}$ is shown in figure 1. Also the energy conditions, $\rho\geq 0$ and $\rho_{p}\geq 0$ are satisfied under conditions
\begin{equation}
\label{eq31}T^{-4}\left[4b+\frac{2\chi}{(2m^{2}+3m+1)}T^{2(2m+1)}\right]\geq\frac{1}{m(m+2)},
\end{equation}
and
\begin{equation}
\label{eq32} T^{-4}\left[-4m(m+2)b+\left(\frac{m^{2}-5m+1}{2m^{2}+3m+1}\right)\chi T^{2(2m+1)}\right]\geq\frac{3}{2},
\end{equation}
respectively. Also $\lambda>0$ under
\begin{equation}
\label{eq33} T>\left[2\left(\frac{2m^{2}+3m+1}{m^{2}-5m+1}\right)\frac{1}{\chi}\right]^{\frac{1}{2(2m-1)}}.
\end{equation}
From Eq. (\ref{eq29}), it is observed that $\lambda$ is an
increasing function of time which is always negative and tends to
zero at late time. It is pointed out by Letelier (1979) that
$\lambda$ may be positive or negative. When $\lambda < 0$, the
string phase of the universe disappears i.e. we have an
anisotropic fluid of particles. This behavior of tension density
$\lambda$ is also
depicted in figure 1.\\

To study the behavior of strings and particles in the universe here we define the following parameter
\begin{equation}
\label{eq34} \frac{\rho_{p}}{|\lambda|}=\frac{-\frac{1}{2}\left(\frac{m^{2}-5m+1}{2m^{2}+3m+1}\right)\chi+m(m+2)bT^{-2(1+2m)}+\frac{3}{4}\frac{1}{T^{2(2m-1)}}}
{|\frac{1}{2}\left(\frac{2m^{2}-3m+1}{2m^{2}+3m+1}\right)\chi-\frac{1}{T^{2(2m-1)}}|}.
\end{equation}
\begin{figure}[htbp]
\centering
\includegraphics[width=8cm,height=8cm,angle=0]{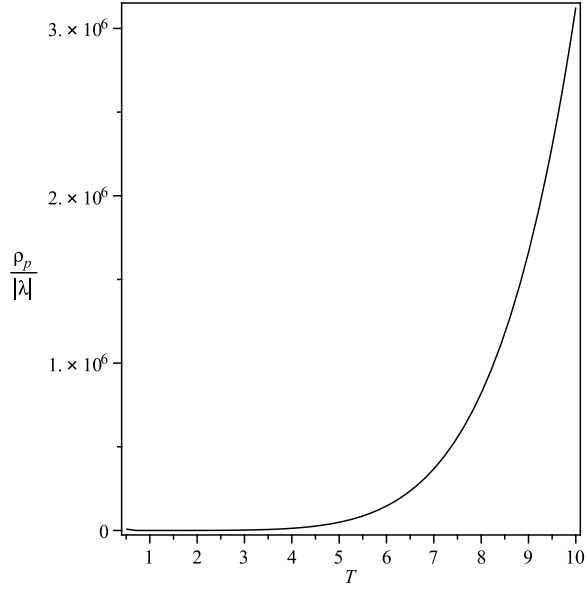}
\caption{The plot of $\frac{\rho_{p}}{|\lambda|}$ Vs. $T$  for $m=2$, $\chi=0.1$ and $b=1$ }
\end{figure}
As mentioned before, since strings are not observed at the present
time of evolution of the universe, in principle we can eliminate
the strings and end up with a cloud of particles. In other word,
we can say the particles dominate over the strings at the present
time of the evolution of the universe. Figure 2 is clearly shows
that in the universe which is describe by the model (\ref{eq27})
the strings dominate over the particle at initial time whereas the
particles dominate over the strings at late time. Also it is worth
to mention that from figures 2 and 3 we observe that at the
initial time, when the universe is in the decelerating phase, the
strings dominate over the particles ($\rho_{p}<\lambda$) whereas
when the universe is in the accelerating phase particles dominate
over the strings ($\rho_{p}>\lambda$). This is in agreement with
the results obtained in Ref (Weinberg. 1976) and (Belinchon. 2009)
and also with the astronomical observations
which predict that there is no direct evidence of strings in the present-day universe.\\

The expressions for the scalar of expansion $\theta$, the average generalized Hubble's parameter $H$, magnitude of shear $\sigma^{2}$, proper volume $V$,
deceleration parameter $q$ and the average anisotropy parameter $A_{m}$ for the model (\ref{eq27}) are given by
\begin{equation}
\label{eq35} \theta=3H=(1+2m)\left[\frac{\chi}{2(2m^{2}+3m+1)}+bT^{-2(1+2m)}\right]^{\frac{1}{2}},
\end{equation}
\begin{equation}
\label{eq36} \sigma^{2}=\frac{(m-1)^{2}}{3}\left[\frac{\chi}{2(2m^{2}+3m+1)}+bT^{-2(1+2m)}\right],
\end{equation}
\begin{equation}
\label{eq37} V=T^{2m+1},
\end{equation}
\begin{equation}
\label{eq38} q=\frac{3}{2m+1}\left[\frac{2bT^{-2(1+2m)}-\frac{\chi}{2(2m^{2}+3m+1)}}{bT^{-2(1+2m)}+\frac{\chi}{2(2m^{2}+3m+1)}}\right],
\end{equation}
\begin{figure}[htbp]
\centering
\includegraphics[width=8cm,height=8cm,angle=0]{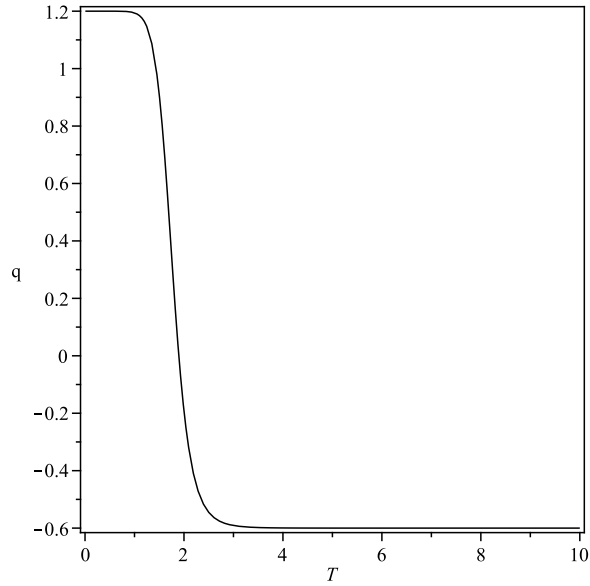}
\caption{The plot of deceleration parameter $q$ Vs. $T$  for $m=2$, $\chi=0.1$ and $b=1$ }
\end{figure}
\begin{equation}
\label{eq39}A_{m}=2\left(\frac{1-m}{1+2m}\right)^{2}.
\end{equation}
From (\ref{eq37}) we observe that
\begin{equation}
\label{eq40}q>0~~if~m>-\frac{1}{2}~\&~T<\left(\frac{2b}{a}\right)^{\frac{1}{2(1+2m)}}~~or~~m<-\frac{1}{2}~\&~T>\left(\frac{2b}{a}\right)^{\frac{1}{2(1+2m)}},
\end{equation}
and
\begin{equation}
\label{eq41}q<0~~if~m>-\frac{1}{2}~\&~T>\left(\frac{2b}{a}\right)^{\frac{1}{2(1+2m)}}~~or~~m<-\frac{1}{2}~\&~T<\left(\frac{2b}{a}\right)^{\frac{1}{2(1+2m)}}.
\end{equation}
A positive sign of $q$ corresponds to the standard decelerating model whereas the negative sign $-1\leq q <0$  indicates the inflation. Recent observations show
that the deceleration parameter of the universe is in the range $-1\leq q <0$ and the present day universe is undergoing an accelerated expansion.
 From figure 3 we observe that the model (\ref{eq27}) successfully describes the expansion of our universe from decelerating to accelerating phase.\\
Also we note that for
\[
T=\left[\frac{(1-m)\chi}{b(2m+7)(2m^{2}+3m+1)}\right]^{-\frac{1}{2(1+2m)}},
\]
$q=-1$ as in the case of de-Sitter universe.\\

In absence of any curvature, matter energy density ($\Omega_{m}$)
and dark energy ($\Omega_{\Lambda}$) are related by the equation
\begin{equation}
\label{eq42} \Omega_{m}+\Omega_{\Lambda}=1,
\end{equation}
where $\Omega_{m}=\frac{\rho}{3H^{2}}$ and $\Omega_{\Lambda}=\frac{\Lambda}{3H^{2}}$.
Thus equation (\ref{eq42}), reduce to
\begin{equation}
\label{eq43}\frac{\rho}{3H^{2}}+\frac{\Lambda}{3H^{2}}=1.
\end{equation}
Using equations (\ref{eq28}) and (\ref{eq35}), in equation (\ref{eq43}), the cosmological constant is
obtained as
\begin{figure}[htbp]
\centering
\includegraphics[width=8cm,height=8cm,angle=0]{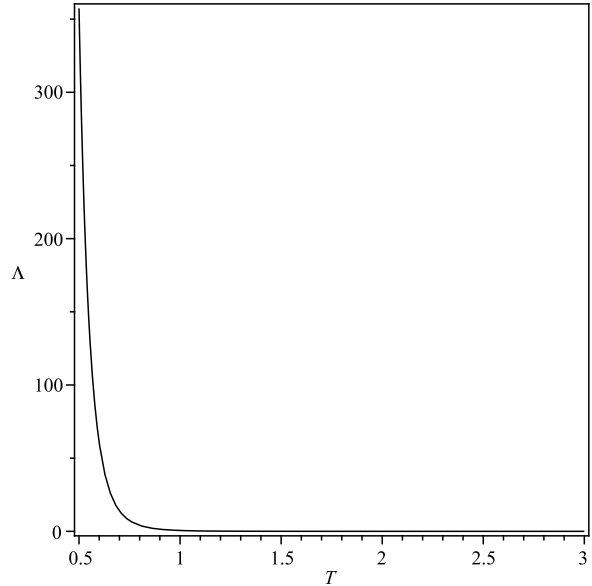}
\caption{The plot of cosmological constant $\Lambda$ Vs. $T$  for $m=2$, $\chi=0.1$ and $b=1$ }
\end{figure}
\begin{equation}
\label{eq44}\Lambda=\left[\frac{(m-1)^{2}}{2(2m^{2}+3m+1)}\chi\right]+\frac{(m-1)^{2}}{3}bT^{-2(1+2m)}+\frac{1}{4}\frac{1}{T^{2(2m-1)}}.
\end{equation}
From (\ref{eq44}) we observe that $\Lambda$ is a decreasing function of time and is always positive for $m>-0.5$ and $m<-1$. This behavior of cosmological constant $\Lambda$ is clearly depicted in figure 4. Recent cosmological observations  suggest the existence of a positive cosmological constant
$\Lambda$ with the magnitude $\Lambda(G\hbar/c^{3})\approx 10^{-123}$. These observations on magnitude and
red-shift of type Ia supernova suggest that our universe may be an accelerating one with induced cosmological
density through the cosmological $\Lambda$-term. Thus, our model is consistent with the results of recent
observations.\\
It is worth to mention that for $m=1$ from (\ref{eq44}) we find
\begin{equation}
\label{eq45}\Lambda=\frac{1}{4}\frac{1}{T^{2}}.
\end{equation}
This supports the views in favor of the dependence
$\Lambda\varpropto T^{-2}$ first expressed by Bertolami (1986 a,
b) and later on observed by several authors (Rahman. 1990; Chen
and Wu. 1990; Berman. 1990 a, b; Berman and Som. 1990; Pradhan and
Kumar 2001). A relation like equation (\ref{eq45}) also can be
found in Brans-Dicke theories when one supposes variable
gravitational
 and cosmological ``constant" (Peebles and Ratra. 2003; Carmeli and Kuzmenko. 2002; Gasperini. 1987).
We have derived the same variation of $\Lambda$ with time in
string viscous cosmology in this article.\\

From the above results, it can be seen that the spatial volume is zero at $T=0$ and it increases with the increase
of $T$ . This shows that the universe starts evolving with zero volume at $T = 0$ and expands with cosmic time $T$ .
In derived model, The energy density $\rho$, particle density $\rho_{p}$, tension density $\lambda$ and the cosmological constant $\Lambda$ become zero at $T \to \infty$
 and tend to infinity at $T = 0$. The model has the point-type singularity at $T = 0$ (MacCallum. 1971).
The expansion scalar and shear scalar all tend to zero as $T\to \infty$. The mean anisotropy parameter is
uniform throughout whole expansion of the universe when $m\ne -\frac{1}{2}$ but for $m= -\frac{1}{2}$ it tends to infinity. This shows that
 the universe is expanding with the increase of cosmic time but the rate of expansion and shear scalar decrease to zero and tend to isotropic.
Since $\frac{\sigma}{\theta}=$constant provided $m\ne
-\frac{1}{2}$, the model does not approach isotropy at any time.
But for $m=1$ our solution provide a totally isotropic universe.

\subsection{Solutions in Absence of Bulk Viscous}
In absence of bulk viscosity, i.e. $\chi\to 0$ or $a\to 0$, the metric (\ref{eq26}) reduces to
\begin{equation}
\label{eq46}ds^{2}=-\frac{dT^{2}}{bT^{-4m}}+T^{2}(dx+zdy)^{2}+T^{2m}(dy^{2}+dz^{2}).
\end{equation}
The pressure $(p)$, the energy density $(\rho)$, the string tension $(\lambda)$, and the particle density $(\rho_{p})$
for the model (\ref{eq46}) are given by
\begin{equation}
\label{eq47} p=\rho=m(m+2)bT^{-2(1+2m)}-\frac{1}{4}\frac{1}{T^{2(2m-1)}},
\end{equation}
\begin{equation}
\label{eq48} \lambda=-\frac{1}{T^{2(2m-1)}},
\end{equation}
\begin{equation}
\label{eq49}\rho_{p}=m(m+2)bT^{-2(1+2m)}+\frac{3}{4}\frac{1}{T^{2(2m-1)}}.
\end{equation}
\begin{figure}[htbp]
\centering
\includegraphics[width=8cm,height=8cm,angle=0]{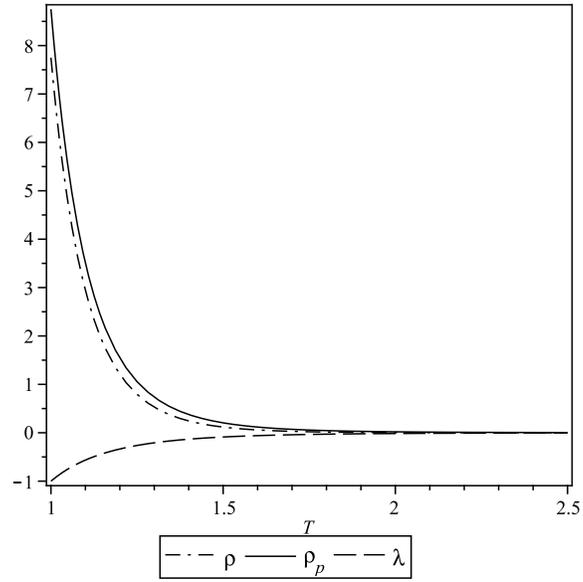}
\caption{The plot of energy density $\rho$, particle density $\rho_{p}$ and tension density $\lambda$ Vs. $T$  for $m=2$ and $b=1$ }
\end{figure}
From eqs. (\ref{eq47}) - (\ref{eq49}) we observe that $\rho$, $\lambda$ and $\rho_{p}$ are decreasing functions of time and $\lambda$ is always negative. The energy conditions, $\rho\geq 0$ and $\rho_{p}\geq 0$ are satisfied under
\begin{equation}
\label{eq50} T\geq \left(4bm(m+2)\right)^{\frac{1}{4}},
\end{equation}
and
\begin{equation}
\label{eq51}m>0~\& m>-2~~or~m<0~\& m<-2.
\end{equation}
respectively. The behavior of $\rho$, $\lambda$ and $\rho_{p}$ are clearly depicted in Figures 5 as a representative
case with appropriate choice of constants of integration and other physical parameters using reasonably well known situations.\\

From eqs. (\ref{eq48}) and (\ref{eq49}) we obtain
\begin{equation}
\label{eq52}\frac{\rho_{p}}{|\lambda|}=bm(m+2)T^{-4}+3/4.
\end{equation}
\begin{figure}[htbp]
\centering
\includegraphics[width=8cm,height=8cm,angle=0]{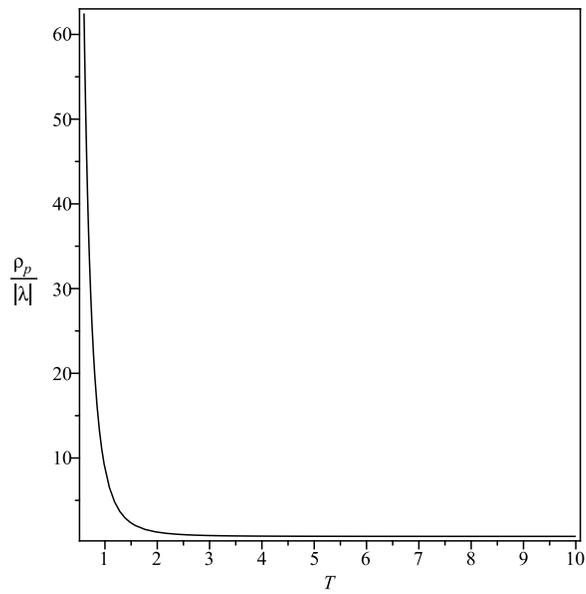}
\caption{The plot of $\frac{\rho_{p}}{|\lambda|}$ Vs. $T$  for $m=2$ and $b=1$ }
\end{figure}
Equation (\ref{eq52}) obviously shows that
$\frac{\rho_{p}}{|\lambda|}$ is a decreasing function of time i.e
as time goes on, the strings dominate over the particles which in
contradict with the result obtained in the first case in presence
of bulk viscosity. This result is of course is not consistent with
the astronomical observations, which predict that there is no
direct evidence of strings in the present-day universe.
Therefore, we conclude that the bulk viscosity may plays an important rule in creation of the particles from strings.\\

The expressions for the scalar of expansion $\theta$, the average generalized Hubble's parameter $H$, magnitude of shear $\sigma^{2}$, proper volume $V$, deceleration parameter $q$ and the average anisotropy parameter $A_{m}$ for the model (\ref{eq46}) are given by
\begin{equation}
\label{eq53} \theta=3H=(1+2m)\sqrt{b}T^{-(1+2m)},
\end{equation}
\begin{equation}
\label{eq54} \sigma^{2}=\frac{(m-1)^{2}}{3}bT^{-2(1+2m)},
\end{equation}
\begin{equation}
\label{eq55} V=T^{2m+1},
\end{equation}
\begin{equation}
\label{eq56} q=\frac{6}{2m+1},
\end{equation}
\begin{equation}
\label{eq57}A_{m}=2\left(\frac{1-m}{1+2m}\right)^{2}.
\end{equation}
From (\ref{eq56}) we observe that $q>0$ if $m>-\frac{1}{2}$ and
$q<0$ if $m<-\frac{1}{2}$. But from eq. (\ref{eq55}) we observe
that, $m<-\frac{1}{2}$ represents an accelerating collapsing
universe with high blue shift. Since the recent observations
(Riess et al. 2001) indicate that we live in an accelerating
expanding universe with red shift, we conclude that in absence of
bulk viscosity a universe with decreasing rate of expansion is the
only possible scenario.

\section {Dark Energy Interpretation of The Models}
Fig. $3$ clearly shows that the presence of bulk viscosity in the
cosmic fluid causes a decelerating to accelerating expansion of
the universe. Also from Raychaudhuri's equation. (\ref{eq17}), we
observe that that bulk viscosity can play the role of an agent
that derive the present acceleration of the Universe.\\
In Eckart's theory (1940) a viscous pressure is specified by
\begin{equation}
\label{eq58} {p}^{eff}= p + \Pi.
\end{equation}
Here $\Pi = -\xi\theta$ is the viscous pressure. Therefore in our
models the effective pressure (stiff fluid plus viscous fluid) can
be written as
\begin{equation}
\label{eq59}p^{eff}=p-\chi=\frac{m(m+2)}{2}\left[\frac{\chi}{(2m^{2}+3m+1)}+2bT^{-2(1+2m)}\right]-\frac{1}{4}\frac{1}{T^{2(2m-1)}}-\chi.
\end{equation}
Using eqs. (\ref{eq28}) and (\ref{eq59}) the effective equation of
state of the net fluid is obtained as
\begin{equation}
\label{eq60}\omega^{eff}=\frac{p^{eff}}{\rho}=1-\frac{\chi}{\frac{m(m+2)}{2}\left[\frac{\chi}{(2m^{2}+3m+1)}+2bT^{-2(1+2m)}\right]-\frac{1}{4}\frac{1}{T^{2(2m-1)}}}.
\end{equation}
\begin{figure}[htbp]
\centering
\includegraphics[width=8cm,height=8cm,angle=0]{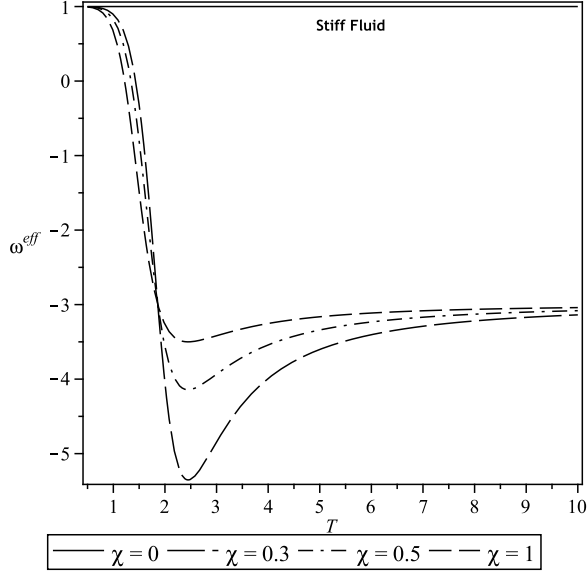}
\caption{The plot of effective equation of state $\omega^{eff}$
Vs. $T$ for $m=1$ and $b=1$ }
\end{figure}
The behavior of effective equation of state, $\omega^{eff}$, in
terms of cosmic time $T$ is shown in Fig. $7$. It is observed that
the $\omega^{eff}$ parameter is a decreasing function of $T$ and
the rapidity of its decrease depends on the value of $\chi$. We
see that in absence of bulk viscosity the models do not exhibit
accelerating expansion (solid line) whereas in presence of
viscosity our models exhibit a decelerating to an accelerating
expansion. From both eq. (\ref{eq60})  and Fig. 7 we observe that
at the later stage of evolution, effective equation of state tends
to the same constant value i.e
$\omega^{eff}=1-\frac{2(2m^{2}+3m+1)}{m(m+2)}$ independent of the
value of $\chi$.

\section {Concluding Remarks}
In this paper we have presented a new exact solution of Einstein's
field equations for LRS Bianchi type II space-time with cloud of
strings which is different from the other author's solution. In
general the models are expanding, shearing and non-rotating. It is
found that in present of bulk viscosity particles dominate over
the strings at late time whereas in absence of viscosity strings
dominate over the particles which is a contradictory result. In
other hand, from Raychaudhuri's equation (\ref{eq17}) we observe
that that bulk viscosity can play the role of an agent that derive
the present acceleration of the Universe. Hence we conclude that
the bulk viscosity plays an important role on the evolution of the
universe. For a universe which was decelerating in past and
accelerating at the present time, the deceleration parameter must
show signature flipping (see the Refs. Padmanabhan and
Roychowdhury. 2003; Amendola. 2003; Caldwell et al. 2006). Our
models are evolving from an early decelerating phase to a late
time accelerating phase (see, Figure 3) which is in good agreement
with recent observations (Riess et al. 2001).
\section*{Acknowledgements}
This work has been supported by the research fund by Mahshahr
Branch of Islamic Azad University under the project entitled
``Study of the homogenous and anisotropic cosmological models by
considering the gravitational effects of viscosity and cosmic
strings". The author also thanks the anonymous referee for the
fruitful comments and suggestions.

\end{document}